\documentclass[preprint,12pt]{elsarticle}
\pdfoutput=1 

\usepackage{amssymb,url,hyperref}
\usepackage{amsmath} 

\begin{document}

\begin{frontmatter}



\title{Hong-Ou-Mandel Interference with a Coexisting Clock using Transceivers for Synchronization over Deployed Fiber}


\author[DSL,NU]{Anirudh Ramesh}
\author[NuCrypt]{Daniel R. Reilly}
\author[NU,NuCrypt]{Kim Fook Lee}
\author[NuCrypt]{Paul M. Moraw}

\author[DSL]{Joaquin Chung}
\author[DSL]{Md Shariful Islam}
\author[Fermi]{Cristián Peña}
\author[CNM]{Xu Han}

\author[DSL]{Rajkumar Kettimuthu}
\author[NU]{Prem Kumar}
\author[NU,NuCrypt]{Gregory Kanter}

\affiliation[DSL]{organization={Data Science and Learning Division, Argonne National Laboratory},
            addressline={9700 S. Cass Avenue}, 
            city={Lemont},
            postcode={60439}, 
            state={IL},
            country={USA}}
\affiliation[NU]{organization={Center for Photonic Communication and Computing, Department of Electrical and Computer Engineering, Northwestern University},
            addressline={2145 Sheridan Road}, 
            city={Evanston},
            postcode={60208}, 
            state={IL},
            country={USA}}
\affiliation[NuCrypt]{organization={NuCrypt LLC},
            addressline={1460 Renaissance Drive, \#205}, 
            city={Park Ridge},
            postcode={60068}, 
            state={IL},
            country={USA}}

\affiliation[Fermi]{organization={Fermi National Accelerator Laboratory},
            addressline={Kirk Road and Pine Street}, 
            city={Batavia},
            postcode={60510}, 
            state={IL},
            country={USA}}

\affiliation[CNM]{organization={Center for Nanoscale Materials, Argonne National Laboratory},
            addressline={9700 S. Cass Avenue}, 
            city={Lemont},
            postcode={60439}, 
            state={IL},
            country={USA}}         

\begin{abstract}

Interference between independently generated photons is a key step towards distributing entanglement over long distances, but it requires synchronization between the distantly-located photon sources.
Synchronizing the clocks of such photon sources using coexisting two-way classical optical communications over the same fiber that transport the quantum photonic signals
is a promising approach for achieving photon-photon interference over long distances, enabling entanglement distribution for quantum networking using the deployed fiber infrastructure.
Here, we demonstrate photon-photon interference by observing the Hong-Ou-Mandel dip between two distantly-located sources: a weak coherent state source obtained by attenuating the output of a laser and a heralded single-photon source. We achieve a maximum dip visibility of $0.58 \pm 0.04$ when the two sources are connected via $4.3$ km of deployed fiber. Dip visibilities $>0.5$ are nonclassical and a first step towards achieving teleportation over the deployed fiber infrastructure.
In our experiment, the classical optical communication is achieved with $-21$ dBm of optical
signal launch power, which is used to synchronize the clocks in the two independent, distantly-located photon sources. 
The impact of spontaneous Raman scattering from the classical optical signals is mitigated by appropriate choice of the quantum and classical channel wavelengths.
All equipment used in our experiment (the photon sources and the synchronization setup)
is commercially available.
Finally, our experiment represents a scalable approach to enabling practical quantum networking with commercial equipment and coexistence with classical communications in optical fiber.
\end{abstract}

\begin{keyword}
Quantum networking \sep Quantum Interference \sep Quantum Optics \sep Single Photons \sep Classical Communications \sep Clock Distribution

\end{keyword}

\end{frontmatter}


\section{Introduction}
\label{sec:sample1}

Photonics is an actively pursued platform for quantum communications due to the ability of photons to be transmitted over low loss optical fiber in the telecommunications band\cite{photonic_quantum_review}.
Quantum networking aims to build infrastructure that transmits photonic quantum states over optical fiber. This can enable future quantum computers and sensors to operate in parallel and elevate their capabilities by increasing their overall computational power, analogous to a high performance computing cluster\cite{distributed_quantum_computing, ionq}.
Significant demonstrations of quantum networking have been made across the world in the United States\cite{barqnet,dcqnet,ieqnet}, UK\cite{UK_quantum}, China\cite{china_qnet} and the European Union\cite{eu_qent}, involving the distribution and measurement of entangled photonic qubits and quantum keys.
However, transmission of entanglement over larger distances is sensitive to attenuation in optical fiber and is bottlenecked by the no-cloning theorem that prevents amplification of quantum states\cite{no_cloning}.
Entanglement swapping with quantum repeaters can help overcome these issues by extending the range of entanglement distribution\cite{repeaters}.
Quantum repeaters can also enable distributed quantum sensing, such as long baseline interferometry for telescopes\cite{sensing} as well as large scale quantum key distribution networks without trusted nodes\cite{qkd_swapping}.
A key step in entanglement swapping using photons involves Hong-Ou-Mandel (HOM) interference of identical photons at a beamsplitter as a part of a Bell state measurement (BSM)\cite{bsm, hom}.
HOM interference involves multiple challenges, for instance alignment of photon polarizations and  arrival times, as well as matching of their spectro-temporal shapes at the beamsplitter.\cite{hom_review}.
Although most examples of HOM interference have been confined to those between two single photon Fock states, it can also occur between a weak coherent state (WCS) and a single photon Fock state\cite{rarity_hom}, with similar experimental constraints. 
A WCS can be obtained by strongly attenuating a laser, to intensities comparable to a single photon Fock state.

Although high visibility HOM interference has been successfully achieved in nearly every photonics platform, it is necessary to achieve reliable HOM interference across fiber networks to connect quantum systems that are not in the same location.
In order to maximize teleportation and swapping rates, one would aim to to use photons confined to a narrow pulse duration to arrive at a beamsplitter for HOM interference, at a timing precision that is within a fraction of their pulse duration\cite{white_rabbit}.
If performed over deployed fiber networks, it becomes necessary to use an optical clock for low latency and high precision clock distribution between the interfering sources to orchestrate their simultaneous arrival at the beamsplitter. 
Using these techniques, several demonstrations of HOM interference have been achieved whether as standalone experiments, or as a part of a BSM for teleporting qubits over deployed fiber networks\cite{hertz_rate,calgary_quantum,teleportation_china,mdiqkd}.
Although most of these experiments achieved high HOM visibilties and successful qubit teleportation/BSM, in most cases the optical clock is transmitted between the nodes in a separate fiber, and there are usually no other signals coexisting with the quantum photonic states.
Some reports have sought to demonstrate coexistence of quantum signals used for HOM/BSM with independent classical traffic\cite{mdiqkd,teleportation_jordan} (with the exception of \cite{coexistence}, which multiplexes a clock signal).

We believe that having the optical clock signals coexisting with the quantum signal in the same fiber is the next step in enabling quantum networks, since fiber infrastructure can factor in as a major overhead cost and fiber availability may be limited.
In a coexisting clock scheme, a common electrical reference is utilized to generate high-precision optical clock signals with modulated lasers. This reference also times the generation of photonic quantum states. Both the clock signals and quantum states are then multiplexed and transmitted through the same fiber\cite{barqnet,picosecond_fermi}.
An advantage of quantum-classical coexistence is that since the properties of optical signals in fiber are sensitive to mechanical and thermal disturbances, a coexisting optical clock signal is sensitive to the same perturbations which affect the quantum signal. 
A coexisting clock with high precision clock recovery has been shown to be achievable\cite{picosecond_fermi} and can significantly improve the accuracy of quantum networking experiments across distant nodes.
However, the optical clock signals can have average optical powers that are in the microwatts to milliwatts range, and hence they can degrade the quantum signal due to spontaneous Raman scattering (SpRS) of the clock signal due to its interaction with the optical fiber\cite{Stolen1984}. The broadband Raman response of silica fiber leads to generation of Raman scattered photons in the quantum signal spectrum\cite{jordan}, which requires careful selection of the quantum and classical channel wavelengths.

In this work, we show successful HOM interference between independently generated WCS and heralded single photons across 4.3 km of deployed underground fiber with a coexisting clock. 
To incorporate scalability in our approach, we interface our commercially available measurement system with inexpensive 2 Gbps small form factor pluggable (SFP) transceivers for clock distribution, and demonstrate a non-classical visibility for HOM interference.

\section{Experimental setup}

HOM interference occurs as a result of destructive interference between the wavefunction amplitudes of two single photons that are sent into the input ports of a beamsplitter, when they are in the same spatio-temporal and polarization modes\cite{hom}.
In HOM interference, coincident detections between the two output ports of a beamsplitter reduce as the photons overlap, which is usually controlled by adding an optical delay line in one of the arms.
The reduction in visibility is a result of both photons exiting the beamsplitter at the same output port, and depends on the indistinguishability between the photons.
Considering $C$ the number of coincident detections in the interferogram, the visibility $V$ is defined in~\cite{hom_review} as follows:
\begin{equation}
    V = \frac{C_{max} - C_{min}}{C_{max}}
\end{equation}
For interference between a WCS and a heralded single photon (HSP) from a photon-pair source, we measured the threefold coincidences between the signal, herald and the WCS to characterize the HOM visibility while varying the delay between the HSP and the WCS\cite{rarity_hom}.
Measurements involving threefold coincidences between a photon-pairs and a WCS have applications in BSM and teleportation experiments\cite{calgary_quantum, hertz_rate} aimed towards enabling entanglement distribution and secure communications\cite{mdiqkd}, and similar HOM measurements  could eventually be used to teleport quantum information.
However, the HOM visibility is limited by multi-pair generation which causes an inherent trade-off between measurement rate (high photon rates) and optimal visibility (low photon rates).
In this work, we generated the HSP using a NuCrypt Entangled Photon Source EPS-1000-W (EPS) \cite{nucrypt_eps1000w}.
The EPS-1000-W is based on a periodically poled lithium niobate (PPLN) waveguide and generates photon-pairs by cascaded second harmonic generation and spontaneous parametric downconversion (c-SHG-SPDC)\cite{Arahira2011}. 
The EPS also has the ability to delay the electrical pulse that drives the intensity modulator that defines its pump.
This can be thus used to control the generation and the subsequent HSP arrival time at the beamsplitter.
In this work, the EPS-1000 pump phase delay has a minimum step size of $10$ ps.
The EPS was pumped with 60 ps pulses centered at 1551.7 nm at a 100 MHz rate, and 100 GHz dense wavelength division multiplexers (DWDMs) were used to demultiplex out photon-pairs from the PPLN c-SHG-SPDC phasematched spectrum at central wavelengths of 1552.89 nm and 1550.2 nm.
The heralding photon at 1552.89 nm was filtered by a 5 GHz tunable fiber Fabry-Perot filter (FFP) after which an ID Quantique single photon detector (SPD) detects the photon to serve as a heralding signal, as shown in Fig. \ref{setup}.
The signal photon was directed towards the HOM interference setup.
The WCS was generated by intensity modulating a Purephotonics Integrable Tunable Laser Assembly (ITLA) using 80 ps electrical pulses, with the same center wavelength as the signal photon, henceforth called the WCS source.
A series of  fixed fiber optic attenuators were used to reduce the probability of a photon per pulse $\bar{n}$ of the WCS source to $\approx 0.01$.
The interfering weak coherent state and idler photon were directed through a fiber polarization controller (FPC), fiber polarizer and a fixed FFP with a 10 GHz bandwidth, to a polarization maintaining (PM) 50:50 coupler for HOM interference, as shown in the schematic in Fig.\ref{setup}. 
The use of a single mode fiber allowed us to enforce spatial indistinguishability, while the PM 50:50 coupler and fiber polarizer helped impose polarization indistinguishability.
The heralding filter and HOM filters had different bandwidths since these were the ones available at the time of the experiment, but typically all the filters would be identical.
Their detection was carried out using two Quantum Opus Superconducting Nanowire Single Photon Detectors (SNSPDs).
The signal, herald and WCS detection signals from the SPAD and SNSPDs were then collected by a NuCrypt Remote Correlation System RCS-2000\cite{nucrypt_rcs2000} (RCS) for correlation measurements. 
The RCS is an electronic correlator which can be used as to generate and recover clock signals and sample detection signals to perform time-correlated measurements.
All the measurements were sampled at a rate of $1.2$ GHz. \\

The EPS and WCS must share a common clock so that their pulse outputs can arrive synchronously at the beam splitter.  
An internal EPS clock serves as the master clock that governs the photon-pair generation. This clock is shared with the WCS by employing a fiber-optic link embedded in the RCS. 
The EPS clock is connected via an SMA electrical cable to a first RCS-2000 system (RCS-EPS). 
The RCS-EPS generates an optical data stream at a nominal 2 Gbps data rate via a SFP transceiver using the EPS clock as a reference clock to an internal phase-locked loop (PLL). 
The optical data stream is connected to a second RCS (RCS-WCS) over a fiber connection. 
The RCS-WCS recovers the clock from the data and generates an output reference clock that is sent to the WCS via an SMA RF cable. 
This experimental topology allowed us to treat the WCS source as an independent node, since it was connected with the rest of the setup with only fiber optic links.
The transceivers used to transmit the optical clock operated at a wavelength of 1310 nm, in the O-band.
The O-band wavelength was chosen due its relatively modest SpRS into the 1550 nm band, low loss in optical fiber, and compatibility with convenient coarse wavelength division multiplexers (CWDMs). 
We note that a 1270 nm wavelength would produce significantly less SpRS noise\cite{jordan} but the RCS-EPS had a 1310 nm SFP installed at the time of this experiment. 

In the next section, we report two experiments that characterized the HOM interference visibility between our WCS and HSP.
In the first experiment, we established a baseline HOM visibility by interfering the HSP and WCS at the beamsplitter.
This experiment demonstrated that we could achieve HOM interference by using an optical clock for synchronizing both interfering quantum signals.
In the second experiment, we implemented the same experiment, except we used a single loop of fiber deployed over the Argonne National Laboratory campus to transmit the WCS state to the beamsplitter, and distribute the optical clock from RCS-EPS to time the WCS generation.
This experiment showed that our quantum signals could still show HOM interference while the WCS was coexisting with a classical signal in the same fiber.
Both the HSP system and the WCS system are highlighted by blue boxes in Fig. \ref{setup}.

\section{Results}
\begin{figure}
    \centering
    \includegraphics[width=1\linewidth]{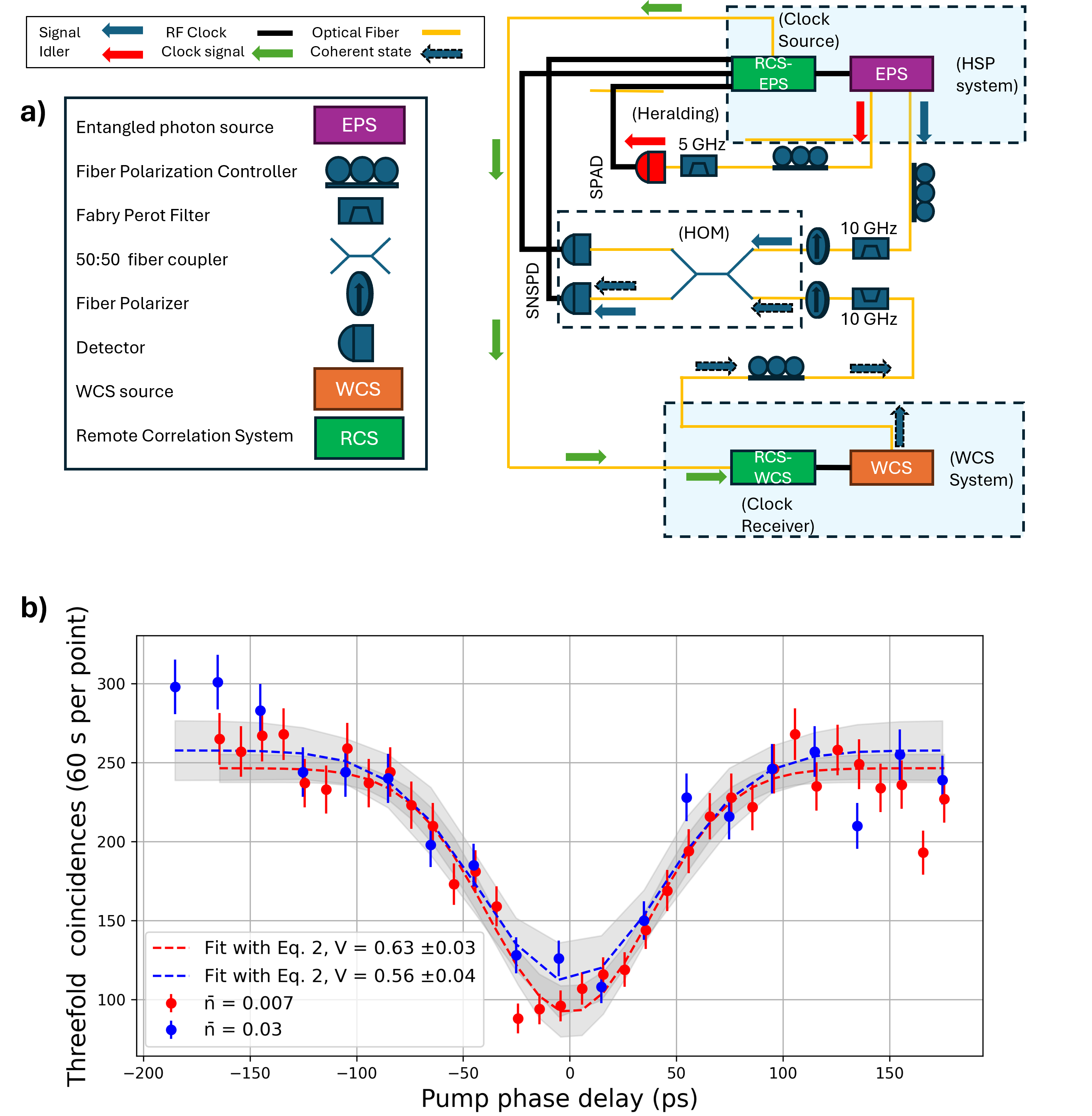}
    \caption{a) Experimental setup for establishing HOM interference baseline, and the b) HOM interferogram. The threefold coincidence counts were integrated at each EPS pump phase delay setting for $60$ s, providing us with a triple coincidences rate of 4.2 triples/second when the pulses were completely misaligned. The error bars on the data are estimated assuming Poissonian distributed arrival times for the photons and the shading represents the 95\% confidence interval for the fit.}
    \label{setup}
\end{figure}

The first experiment sought to set a baseline HOM visibility for this system, by measuring the maximum HOM visibility achievable with the optical components available.
After setting up the system as shown in Fig. \ref{setup}, FPCs were used to align the polarizations of the incoming WCS and HSP with the polarizers before the 50:50 coupler. Polarization indistinguishability is ensured by maximizing the detected counts at the SNSPDs.
Following that, the EPS pump delay was scanned.
This changed the arrival times of the HSP at the beamsplitter, and allows control over the temporal overlap of the HSP and WCS pulses at the 50:50 coupler. 
The acquired interferogram was then fit to the following expression for the threefold coincidences $C_{123}$\cite{hom_formula}:

\begin{equation}
    C_{123} = C\times(1-V\times e^{-(t/\tau)^2})
    \label{fringe}
\end{equation}
This is an inverse Gaussian function where $V$ is the interference visibility, $C$ is the maximum threefolds coincidences and $\tau$ is related to the standard deviation $\sigma$ of the Gaussian by the relation:
$\sigma = 2\tau\times (log_e (2))^\frac{1}{2}$. In the first experiment, the WCS $\bar{n}_{baseline}$ was estimated to be $0.007$ (with $\bar{n}$ defined as the average WCS photon number per pulse the input to the 50/50 coupler) and the coincidences-to-accidentals ratio (CAR) of the photon-pair source was estimated to be $40$. The former was estimated by observing the singles count rate due to the WCS (assuming a typical SNSPD efficiency), and the latter was calculated by measuring correlations between the signal and idler photons in RCS-EPS. 

This experiment resulted in a HOM visibility of $0.63 \pm 0.02$. 
This visibility is unambiguously non-classical, and cannot be attributed to classical interference.
However, the imperfect visibility can be attributed to three main factors.
The first was due to mismatch in the spectro-temporal modes between the HSP and the WCS. 
These modes are defined by a combination of the spectral filters and the input pulse-shapes. 
The 5 GHz filter before the heralding detector set the effective bandwidth of the HSP, and the $\approx80$ ps full-width at half maximum (FWHM) pulse duration of the WCS created an estimated spectral FWHM of $5.5$ GHz for the WCS, assuming a transform limited Gaussian as the pulse shape as a reasonable approximation.
Hence, this mismatch was expected to lower the HOM interference visibility.
A second factor was the timing jitter between the recovered clock at RCS-WCS and the EPS clock, which was measured using a digital communications analyzer (sampling oscilloscope) to be about 20 ps RMS when the systems were connected by a short fiber. 
No further efforts were made to optimize the filters or pulse-shapes for maximum interference since our focus was to study HOM with the coexistence of an optical clock. 
A third factor limiting visibility are multi-photon events from the HSPS or WCS, which are controlled but not eliminated by operating at low photon levels. 
\begin{figure}
    \centering
    \includegraphics[width=1\linewidth]{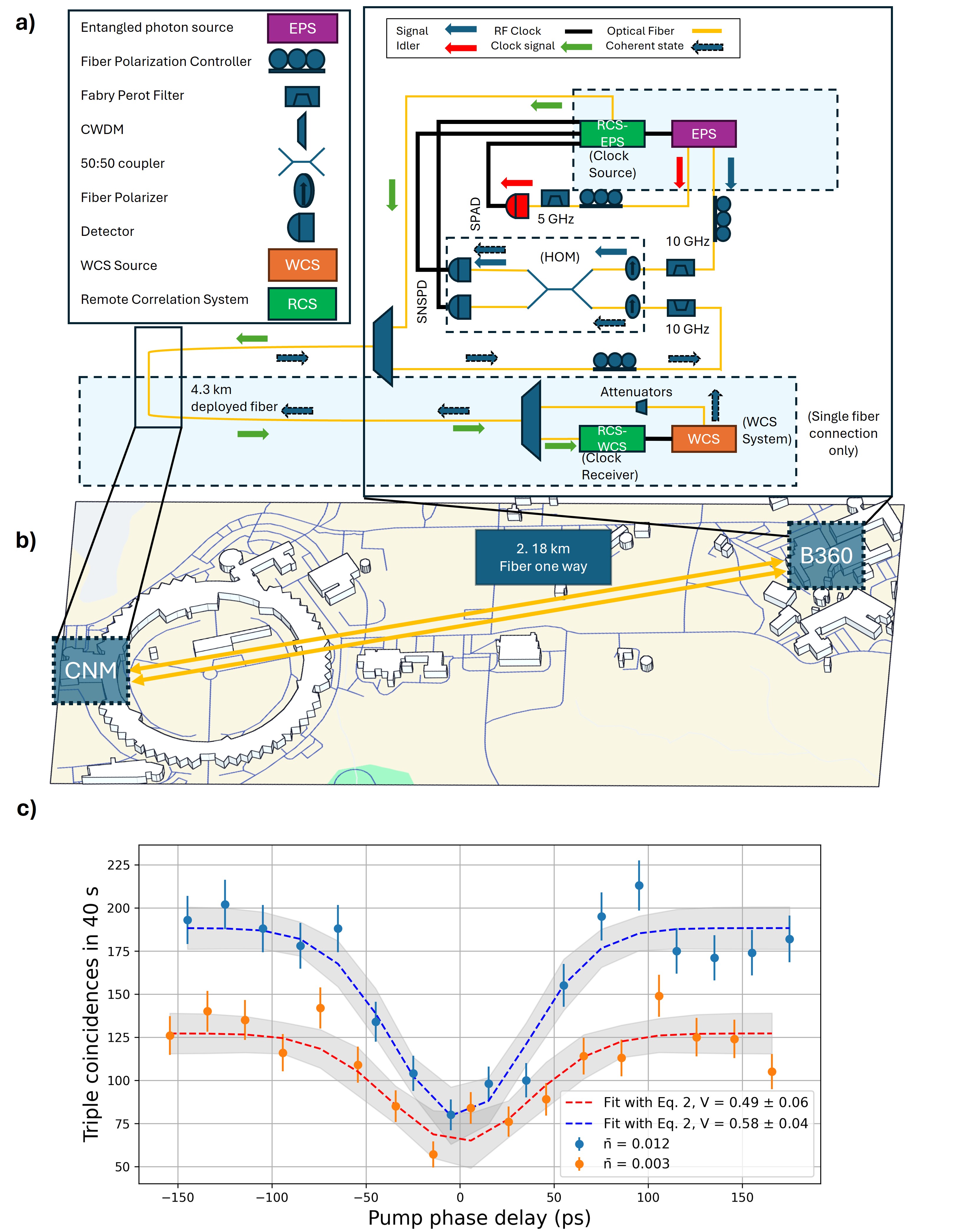}
    \caption{a) Experimental setup, b) map of deployed fiber loop for HOM with quantum-classical coexistence at Argonne National Laboratory, and c) HOM interferograms for $\bar{n}$ = 0.012 and $\bar{n}$ = 0.003. The map was acquired from Cadmapper\cite{cadmapper}.}
    \label{loop}
\end{figure}
\\\\

To test this approach in a networked setting, a pair of deployed optical fibers between the site of the experiment and the Center for Nanoscale Materials building (CNM) were looped back at CNM, as shown in Fig. \ref{loop} b).
A Polatis series 6000 software-defined networking switch was used to route the WCS and clock signals from their sources to the deployed fiber loop.
Optical time-domain reflectometry (OTDR) was used to determine the overall fiber loop length as $4.3$ km. 
A total of 6 dB of loss was measured for the fiber loop in the C-band.
This loss can be attributed to multiple junctions in the fiber network.
As displayed in Fig. \ref{loop}, this experiment sought to emulate two independent and separated quantum nodes both connected by a single strand of deployed fiber.
The fiber loop carried the WCS to the beamsplitter, as well as a coexisting counter-propagating classical clock signal from RCS-EPS which was recovered by the RCS-WCS and used to time the WCS generation.
O-band/C-band CWDMs were used to multiplex and separate the WCS and clock signals.
This enabled the synchronization of the generation of the WCS and the HSP.
In order to reduce SpRS noise from the optical clock, we aimed to have as low a coexisting classical launch power as possible that could be successfully recovered by the transceivers.
Fiber optic attenuators were added at the output of the transceiver that generated the clock signal before being launched into the loop, and a launch power of -21 dBm was measured.
The clock signal was estimated to have an optical power level of $\approx-27$ dBm after travelling over the deployed fiber, which was still successfully recovered by RCS-WCS after being demultiplexed from the fiber loop by the CWDM.
Using these settings, two HOM interferograms were acquired and are shown in Fig. \ref{loop} c). 
Each interferogram in this experiment was acquired with the WCS having a different $\bar{n}$, namely $\bar{n}_{loop-1}$ = $0.012$ and $\bar{n}_{loop-2}$ = $0.003$. 
For the case with  $\bar{n}_{loop-1}$ = $0.012$, we observed a reduced visibility of $0.58 \pm 0.04 \ $  compared to the baseline visibility of $\approx 0.63$ for $\bar{n}_{baseline}$ = $0.007$.
Hence, our coexisting optical clock distribution method helped maintain sufficient timing synchronization between the HSP and WCS systems to measure a non-classical HOM fringe.
For the next measurement, we further reduced $\bar{n}$ by programming a variable optical attenuator in the WCS source while keeping all other parameters the same, to $\bar{n}_{loop-2}$ = $0.003$.
The resulting interferogram brought the visibility into the classical regime, which is defined as V $<0.5$, to V = $0.49 \pm 0.06$.\\

To understand these visibilities, we refer to the following model\cite{Xu2023}:
\begin{equation}
   V_{\text{HOM}} = \frac{4 \delta\omega_A \delta\omega_B}{(\frac{\bar{n}}{\mu} + 2 + \frac{N_{\text{sys}}}{\bar{n} \mu})(\delta\omega_A^2 + \delta\omega_B^2)} \exp\left(-\frac{(\delta\omega_A^2 \delta\omega_B^2) t^2}{\delta\omega_A^2 + \delta\omega_B^2}\right)
   \label{fit}
\end{equation}
Here the measured HOM visibility is described as a function of $\bar{n}$, the probability of a HSP arriving at the beamsplitter per pulse $\mu$,  the spectral bandwidths of the WCS and HSP, $\delta\omega_A$ and $\delta\omega_B$ respectively, and $N_{sys}$, which is the probability of a threefolds coincidence resulting either from noise or higher photon number states from the WCS.

According to this model, we expect an optimal visibility limited by noise from the laser and background counts.
We assume a common noise level for all our experiments, since our choice of wavelengths would greatly reduce SpRS from the clock.
Upon fitting for this expression using the lmfit package in Python with $\delta\omega_A = 11\pi\times 10^9$  rad,  $\delta\omega_B = 10\pi\times 10^9$ rad ($5.5$ and  $5$ GHz in angular units) at $t = 0$ s.
The fit converges, although sub-optimally.
It should be noted that the values for the spectral bandwidths are estimates calculated by the combinations of filters and pulse durations used in the experiment, and not exact values.
Additionally, other experimental effects such as jitter are not included in the analysis. The fitting is thus not expected to be very accurate but is a mechanism to explore trends in the data.

\begin{figure}
    \centering
    \includegraphics[width=1\linewidth]{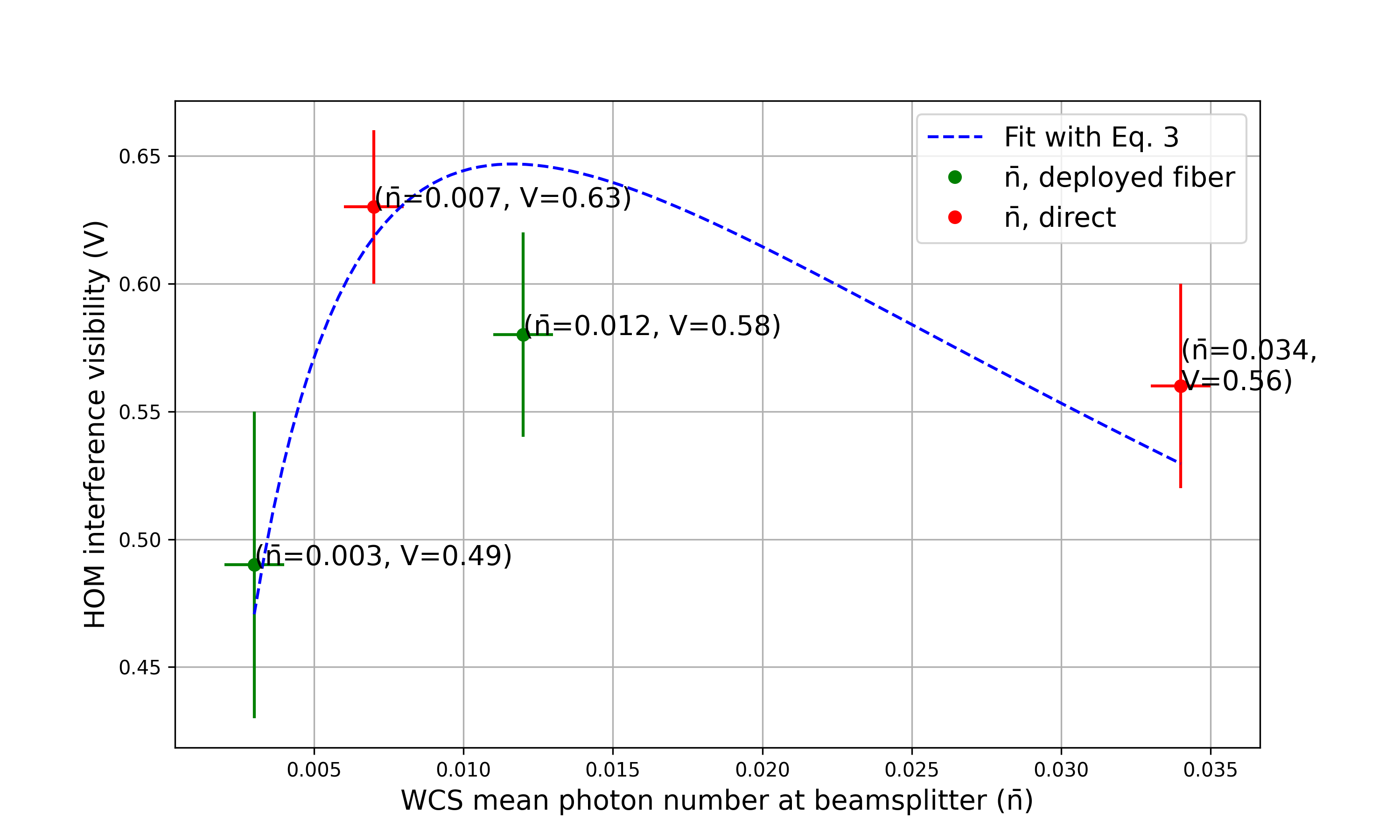}
    \caption{Fit obtained from Eq \ref{fit}. The y-axis error bars for the data points were obtained by fitting the HOM interferograms to Eq \ref{fringe} and the x-axis error bars were based on experimental uncertainties.} 

        \label{fig:Data analysis}
\end{figure}
Based on the fit in Fig. \ref{fig:Data analysis}, we obtain $N_{sys} = 1.3\pm 0.47\sep\times 10^{-4}$  per pulse and $\mu = 0.021 \pm 0.004$ per pulse.
$\mu$  here can be interpreted as the  probability of a photon entering the HOM beamsplitter conditioned on a heralding detection event per pulse.
This can be independently estimated \cite{kwiatmu} as ~$0.04$ using  the relation $P(h|s) = \mu = \frac{P_{cc}}{P_{h}\times\eta_{s}}$  where $P_{cc}$ is the probability of a coincidence between the heralding SPD and both SNSPDs, $P_h$ is the probability of a singles count in the heralding SPAD and $\eta_s$ is the efficiency of one of the the singles channels.
By adding count probabilities from both SNSPD channels, we find $P_{cc} = 8\times10^{-6}$.
$P_h$ was estimated to be $2.6\times 10^{-4}$ and $\eta_s$ was estimated to be $0.72$, which provides with our independent value for $\mu$.
The fit produces a lower heralding efficiency estimate. However, as previously noted, Eq. \ref{fit} is a simple model that is used here primarily for investigating general trends and is not expected to lead to the most accurate parameter estimates. 
The $N_{sys}$ value in our case is expected to be a combination of noise from the EPS, and higher order photon Fock states from WCS and the EPS.

Broadly, our visibility measurements follow a trend similar to that predicted by Eq. \ref{fit}, such that there is an optimal value for the visibility based on the interplay of $N_{sys}$, $\mu$ and $\bar{n}$.
Further detailed characterization efforts will be helpful to pinpoint parameters more accurately and achieve higher quality fits.

\section{Discussion}
In this work, we achieved HOM interference between a WCS and HSP with a coexisting optical clock in the same fiber, which was generated and received by commercial SFP transceivers.
In spite of a mismatch in the spectro-temporal properties of the pulses and classical coexistence, we still observed non-classical HOM visibility.
Although our transceivers had some jitter, we were still able to push the limits of the experiment above $V = 0.5$ (the classical regime), and we expect that the use of faster transceivers can close the gap in the timing synchronization precision between transceivers and sophisticated modulated laser pulse sources.
A better HOM visibility can also be achieved by using matching filter bandwidths for all the channels involved in the measurement.
As quantum networks develop alongside advances in quantum computing and transduction, using existing tools in the telecommunications industry like transceivers to implement entanglement distribution will enable rapid scalability in quantum communications.
Coexisting classical and quantum communications would enable rapid proliferation of quantum traffic by sharing the existing infrastructure.

\appendix

\section{Acknowledgements and funding}
\label{sec:sample:appendix}
We thank Jordan M. Thomas, Northwestern University for useful discussions.
This material is based upon work supported by the U.S. Department of Energy (DOE), Office of Science, Advanced Scientific Computing Research (ASCR) program under contract number DE-AC02-06CH11357 as part of the InterQnet project, and the DOE ASCR Transparent Optical Quantum Networks for Distributed Science program through the Illinois Express Quantum Network project, but no government endorsement is implied. 
Part of this work was also performed at the Center for Nanoscale Materials, a U.S. Department of Energy Office of Science User Facility, was supported by the U.S. DOE, Office of Basic Energy Sciences, under Contract No. DE-AC02-06CH11357.

\bibliographystyle{elsarticle-num}

The submitted manuscript has been created by UChicago Argonne, LLC, Operator of Argonne National Laboratory (``Argonne''). Argonne, a U.S. Department of Energy Office of Science laboratory, is operated under Contract No. DE-AC02-06CH11357. The U.S. Government retains for itself, and others acting on its behalf, a paid-up nonexclusive, irrevocable worldwide license in said article to reproduce, prepare derivative works, distribute copies to the public, and perform publicly and display publicly, by or on behalf of the Government.  The Department of Energy will provide public access to these results of federally sponsored research in accordance with the DOE Public Access Plan. http://energy.gov/downloads/doe-public-access-plan.




\end{document}